\documentclass[12pt]{article}
\usepackage{epsf}
\def\beq{\begin{equation}}
\def\eeq{\end{equation}}

\begin{document}
\begin{titlepage}
\begin{center}
{\Large \bf William I. Fine Theoretical Physics Institute \\
University of Minnesota \\}
\end{center}
\vspace{0.2in}
\begin{flushright}
TPI-MINN-06/28-T \\
UMN-TH-2514-06 \\
August 2006 \\
\end{flushright}
\vspace{0.3in}
\begin{center}
{\Large \bf Possible new resonance at the $D^* {\bar D^*}$ threshold in $e^+e^-$
annihilation
\\}
\vspace{0.2in}
{\bf S. Dubynskiy \\}
School of Physics and Astronomy, University of Minnesota, \\ Minneapolis, MN
55455 \\
and \\
{\bf M.B. Voloshin  \\ }
William I. Fine Theoretical Physics Institute, University of
Minnesota,\\ Minneapolis, MN 55455 \\
and \\
Institute of Theoretical and Experimental Physics, Moscow, 117218
\\[0.2in]
\end{center}

\begin{abstract}
We argue that the recent CLEO-c data on $e^+e^-$ annihilation into pairs of
charmed mesons at c.m. energy around 4.0\,GeV are not well described by a single
resonance $\psi(4040)$, but can be better understood if there is an additional
narrow resonance with mass within few MeV from the $D^* {\bar D}^*$ threshold.
\end{abstract}

\end{titlepage}

The strong dynamics of heavy and light quarks  gives rise to rich structures
near the thresholds of open heavy flavor states. In particular, the cross
section of the $e^+e^-$ annihilation is well known to display an intricate
behavior in the region of the thresholds for production of various pairs of $D$
($D^*$) mesons\cite{pdg}. The most prominent peak in the annihilation cross
section is conventionally labeled\cite{pdg} as the $\psi(4040)$ resonance. It
has been argued long ago\cite{drgg} that this resonance (then called
$\psi(4028)$) might be of  a `molecular'\cite{ov} type, i.e. essentially made of
a meson pair $D^* {\bar D}^*$, on the basis of the observed great enhancement of
its coupling to $D^* {\bar D}^*$  in comparison with its interaction with the
other open charmed meson channels: $D {\bar D}$ and $D {\bar D}^*$. However it
was later pointed out\cite{yopr} that a similar pattern of couplings to meson
pairs can arise for a pure charmonium state due to a kinematical cancellation in
overlap integrals at the momentum corresponding to the meson states $D {\bar D}$
and $D {\bar D}^*$. In other words, the latter scheme suggests a suppression of
these lighter channels rather than an enhancement of the coupling to $D^* {\bar
D}^*$. This issue has not been resolved so far due to lack of more detailed data
on the production of the charmed mesons in  $e^+e^-$ annihilation in the
vicinity of the peak. Recently the CLEO-c experiment has produced\cite{poling} a
greatly improved set of data from the scan of the $e^+e^-$ production cross
section separately for each open channel with charmed meson pairs. The data
display a highly nontrivial behavior in each of the channels in all the scanned
energy range from 3970\`MeV to 4260\,MeV, and in particular at energies around
the $\psi(4040)$ peak. The purpose of this letter is to point out that the CLEO-
c data in the immediate vicinity of the threshold for the $D^*{\bar D}^*$ meson
pairs, specifically in the energy range from 3970\,MeV to 4060\,MeV, cannot be
described with a reasonable statistical significance by a single resonance and a
non-resonant background. Rather we find that the data suggest that in addition
to a resonance with parameters close to those of $\psi(4040)$ there exists a
narrow structure, with a width of not more than few MeV, at a c.m. energy very
close to 4010$\, \div \,$4015\,MeV, i.e. very close to the thresholds for the
neutral vector meson pairs $D^{*0} {\bar D}^{*0}$ ($4013 \pm 1\,$MeV) and for
the charged ones $D^{*+} D^{*-}$ ($4020 \pm 1\,$MeV). The available CLEO-c data
however are yet not detailed enough to fit the parameters of the resonance
suggested by this structure. We were only able to make a very preliminary
estimate of the $e^+e^-$ branching fraction of such resonance as ${\cal B}_{ee}
\sim 10^{-7}$.

If confirmed by a more detailed experimental study the suggested resonance would
tantalizingly invite a `molecular' interpretation and an analogy with the now
well known resonance $X(3872)$\cite{belle} at the threshold of the meson pair
$D^0 {\bar D}^{*0}$. However, unlike the $X(3872)$ resonance, which corresponds
to an $S$-wave meson pair, the suggested resonance would correspond to a
$P$-wave charmed meson pair. The appearance of threshold singularities in
different partial waves can imply a quite nontrivial dynamics of the strong
interaction between heavy-light mesons.

In what follows we use the CLEO-c scan data\cite{poling} for the $e^+e^-$
annihilation cross section at six values of the c.m. energy (in MeV): 3070,
3090, 4010, 4015, 4030 and 4060. The cross section data are available separately
for the channels $D {\bar D}$, $D_s {\bar D}_s$, $D {\bar D}^*$ and $D^* {\bar
D}^*$. Where appropriate, these data are for the sum over the pairs of charged
and neutral mesons, and we assume that the mesons are produced by the
isotopically singlet electromagnetic current of the charmed quarks, and the
isotopic asymmetry, relevant for the channel $D^* {\bar D}^*$, due to the
isotopic mass difference and due to the Coulomb interaction is taken into
account. We attempt a fit of the data to a model of a smooth background and a
resonance, taking into account the interference effects and the effects due to
the onset of the $D^* {\bar D}^*$ channel at the threshold and the channel
mixing. In the actual calculation we use, instead of the cross sections, the
dimensionless production rate coefficients $R_k$ defined as
\begin{eqnarray}
\label{rs}
\sigma(e^+e^- \to D {\bar D}) = \sigma_0(s)\, 2 \, v_D^3 \, R_1~,&~&~
\sigma(e^+e^- \to D_s {\bar D}_s) = \sigma_0(s)\, v_{D_s}^3 \, R_2~,  \\
\nonumber
\sigma(e^+e^- \to D {\bar D}^*+D^* {\bar D}) = \sigma_0(s)\, 6 \, \left ({2 p}
\over \sqrt{s} \right )^3 \, R_3~,&~&~\sigma(e^+e^- \to D^* {\bar D}^*) =
\sigma_0(s)\, 7 \, (v_0^3+v_+^3) \, R_4~,
\end{eqnarray}
with $\sigma_0=\pi \, \alpha^2/(3 s)$. Here $v_D$, $v_{D_s}$, $v_0$ and $v_+$
stand for the c.m. velocities of each of the mesons in respectively the channels
$ D {\bar D}$, $D_s {\bar D}_s$, $D^{*0} {\bar D}^{*0}$ and $D^{*+} {\bar
D}^{*-}$, while for the channel $D {\bar D}^*+D^* {\bar D}$ with mesons of
unequal mass the velocity factor is replaced by $(2 p/\sqrt{s})$ with $p$ being
the c.m. momentum. The extra factors 1, 3 and 7 in respectively the
pseudoscalar-pseudoscalar, pseudoscalar-vector and the vector-vector channels
correspond to the ratio of the corresponding production cross section in the
simplest model\cite{drgg,cornell} of independent quark spins, so that the
inequality between $R_1$, $R_3$ and $R_4$ also illustrates a conspicuous
deviation from this model.

\begin{table}[t]
\caption{The dimensionless production rate
coefficients $R_k$ defined by Eq.(\ref{rs}) calculated from the CLEO-c scan
data\cite{poling}.}
\label{tab1}
\begin{tabular}{|c|cccccc|}
\hline
E({\rm MeV}) & 3070 & 3090 & 4010 & 4015 & 4030 & 4060 \\
\hline
$R_1$ & $1.89 \pm 0.27$ & $1.18 \pm 0.23$ & $0.61 \pm 0.15$ & $0.28 \pm 0.24$ &
$2.92 \pm 0.27$ & $3.69 \pm 0.25$ \\
$R_2$ & $34.2 \pm 8.7$ & $22.6 \pm 5.3$ &  $28.9 \pm 3.2$ & $23.0 \pm 5.4$ &
$14.0 \pm 2.9$ &  $2.7 \pm 1.5$ \\
$R_3$ & $49.8 \pm 1.4$ & $45.6 \pm 1.2$ & $45.4 \pm 0.8$ & $47.9 \pm 1.6$ &
$36.8 \pm 0.9$ & $17.5 \pm 0.6$ \\
$R_4$ & --- & --- & --- & $709 \pm 161$ & $357 \pm 14$ & $81.8 \pm 3.1$ \\
\hline
\end{tabular}
\end{table}

The values of the coefficients $R_i$ corresponding to the CLEO-c data at the
discussed energies are given in the Table~\ref{tab1}. One can readily see from
the table that the rate coefficient $R_4$ is very large near the threshold of
the $D^* {\bar D}^*$ pairs, a behavior noticed long ago\cite{drgg}, and which
behavior can make sense only if the production of the vector meson pairs is
dominated by a resonance, while for the other channels both a resonant and a
non-resonant production amplitudes are generally present. Accordingly, we
parametrize the rate coefficients as\cite{mv0}
\beq
R_k(E) = \left | a_k + {b_k + {\rm i} \, c_k \over D(E)} \right |^2
\label{r13}
\eeq
for k=1, 2, and 3, and
\beq
R_4(E)={b^2 \over |D(E)|^2}~,
\label{r4}
\eeq
where $a_k$, $b_k$,  $c_k$, and $b$ are real numbers\,\footnote{The
insensitivity of the cross section to the overall phases of the production
amplitudes allows one to set e.g. the fit parameters $a_k$ and the factor $b$
for the $D^* {\bar D}^*$ channel as real.},  and $D(E)$ is the resonant
denominator, which takes into account the threshold effects due to the strong
coupling of the resonance to the $D^* {\bar D}^*$ channel. The expression for
the latter factor at $E$ above the $D^{*+} D^{*-}$ threshold has the form
\beq
D=E-W_0 + {{\rm i} \over 4}\, \left [ 2 \, \Gamma_0 + {(E-2 \, M_0)^{3/2} \over
w^{1/2} }+ {(E-2 \, M_+)^{3/2} \over w^{1/2} } \right ]~,
\label{dh}
\eeq
where $M_0$ ($M_+$) is the mass of the $D^{*0}$ ($D^{*+}$) meson, $W_0$ is the
`nominal' mass of the resonance, and $\Gamma_0$ is the width of the resonance
decay into all channels except for $D^* {\bar D}^*$. The decay rate into the
latter final states rapidly changes near the threshold and is accounted by the
last two terms in the square braces in Eq.(\ref{dh}) with $w$ being a parameter
with dimension of energy, describing the coupling of the discussed resonance to
$D^* {\bar D}^*$. At energy $E$ below one or both of the $D^* {\bar D}^*$
thresholds, the resonance denominator $D(E)$ is found from Eq.(\ref{dh}) by
analytical continuation. Thus at the energy between the thresholds for the
neutral and the charged $D^*$ mesons, $2 \, M_0 < E < 2 \, M_+$, one finds
\beq
D={1 \over 4} \, {(2 \, M_+ - E)^{3/2} \over w^{1/2} }+E-W_0 + {{\rm i} \over
4}\, \left [ 2 \, \Gamma_0 + {(E-2 \, M_0)^{3/2} \over w^{1/2} } \right ]~,
\label{dm}
\eeq
and below both thresholds, i.e. at $E < 2 \, M_0$,
\beq
D={1 \over 4} \,\left[ {(2 \, M_+ - E)^{3/2} \over w^{1/2} }+{(2 \, M_0-E)^{3/2}
\over w^{1/2} } \right ]+E-W_0 + {{\rm i} \over 2} \, \Gamma_0  ~.
\label{dl}
\eeq

Using the described model of a resonance plus background production of the
charmed mesons we have made a fit to the data in Table~\ref{tab1}. Although the
resulting central values of the fit parameters ($W_0=4023\,$MeV,
$\Gamma_0=67\,$MeV) for the resonance are in a reasonable agreement with the
data listed in the PDG Tables\cite{pdg} for $\psi(4040)$, the statistical
significance of the fit is quite poor: $\chi^2/N_{\rm dof}=17.6/8$. We also have
not found any improvement in the statistical significance by introducing a
linear in the momentum form-factor $(1-p^2/\mu^2)$ in the amplitudes. Another
modification of the described model by introducing a non-resonant background for
the $D^*$ pair production amplitude is not technically feasible right now due to
lack of data. Indeed, there are in fact two $P$-wave amplitudes describing the
production of the $D^* {\bar D}^*$ pairs near the threshold: one with the total
spin of the meson pair $S=0$ and the other with $S=2$\,\footnote{It looks
reasonable to neglect in the threshold region the third possible amplitude,
corresponding to an $F$-wave production with $S=2$.}. The resonant production
amplitude in the considered model corresponds to a fixed composition of these
two amplitudes, which composition is generally different in the non-resonant
part, which thus introduces at least four new parameters. It should be however
noticed that introducing a non-resonant production of the $D^*$ mesons at the
threshold is neither physically palatable, nor it is likely to improve agreement
with the data. Indeed, in order to significantly affect the cross section the
non-resonant amplitude should be comparable with the resonant one, which is very
big, as can be seen from the $R_4$ entries in Table~\ref{tab1}. We are not aware
of any reasonable mechanism that would explain such an enhancement of the non-
resonant amplitude. On the other hand the rather large width of the $\psi(4040)$
resonance implies only a smooth variation of the cross section on the energy
scale considered here. The data however apparently display a small but sharp
`wiggle' in the $D {\bar D}$ data at 4010 and 4015\,MeV and a possible steep
rise towards the threshold in the $D^* {\bar D}^*$ cross section at 4015\,MeV.
In fact this feature in the data is the main contributor to a large $\chi^2$ in
our fit. Thus introducing a smooth non-resonant term in the $D^* {\bar D}^*$
production amplitude is very unlikely to improve the situation.

An obvious effect, which is significantly enhanced in the immediate vicinity of
the threshold for the charged mesons, $D^{*+}D^{*-}$, is the Coulomb interaction
between the mesons. Therefore one can probe this effect as a possible reason for
the steeper behavior of the cross sections near the threshold indicated by the
data. However, we have not found any improvement in the statistical significance
of the fit by including the Coulomb interaction. In other words at the available
scan points the Coulomb effect is too weak to be visible in the data. We
nevertheless present here the relevant formulas, which can be helpful if more
detailed data immediately near the threshold become available.

Generally, the effect of the Coulomb interaction can be
complicated\cite{mv} in the resonance region by the interference with the strong
phase. However due to the relatively large width of $\psi(4040)$ the variation
of the strong phase on the scale of few MeV around the threshold (where the
Coulomb effect is most essential) can be neglected, and one can neglect such
interference and estimate the effect of the Coulomb interaction by considering
the slow $D^{*\pm}$ mesons as point charges. In this approximation the Coulomb
effect in the $D^{*+}D^{*-}$ channel can be found by adapting the textbook
formulas\cite{ll}. The corrected expression for $R_4$ takes the form
\beq
R_4(E)={b^2 \over |D_c(E)|^2} \left [ {v_0^3 + v_+^3 F_c(E) \over v_0^3 + v_+^3}
\right ] ,
\label{r4c}
\eeq
and in the expressions for rest of the rate coefficients, as well as in the
latter formula, the resonant denominator $D(E)$ is replaced by the
Coulomb-corrected $D_c(E)$, which is described as follows. At $E > 2 \, M_+$,
i.e. above the charged meson threshold the term $(E-2 \, M_+)^{3/2}/ w^{1/2}$ in
Eq.(\ref{dh}) is replaced as
\beq
 {(E-2 \, M_+)^{3/2} \over w^{1/2}} \to {(E-2 \, M_+)^{3/2} \over w^{1/2}}\,
F_c(E)~,
\label{dhc}
\eeq
while below the charged meson threshold the corresponding term in the equations
(\ref{dm}) and (\ref{dl}) is replaced as
\beq
{(2 \, M_+ - E)^{3/2} \over w^{1/2} } \to {(2 \, M_+ - E)^{3/2} \over w^{1/2} }
{\tilde F}_c(E)~,
\label{dmlc}
\eeq
where $F_c$ (${\tilde F}_c$) is the Coulomb factor in the $P$ wave above (below)
the threshold. Above the threshold this factor is conveniently expressed in
terms of the Coulomb parameter $\lambda = M_+ \alpha/(2 k)$, where $k=\sqrt{ M_+
(E-2 M_+)}$ is the c.m. momentum of the $D^*$ meson:
\beq
F_c={2 \pi \, \lambda \, (1+\lambda^2)\over 1-\exp(-2 \pi \, \lambda)}~.
\label{fc}
\eeq
The analytical continuation of this expression below the threshold can be
written in terms of the real quantity $\nu=M_+ \alpha/(2 \kappa)$ with
$\kappa=\sqrt{ M_+ (2 M_+-E)}$:
\beq
{\tilde F_c}= (1-\nu^2) \, \nu \,  \left ( \pi \,  \cot \pi \nu - 2 \, \ln
{\kappa_0 \over \kappa} \right )~.
\label{fct}
\eeq
The parameter $\kappa_0$ in the latter formula can be chosen arbitrarily, and
any variation in such choice can be absorbed in a shift in $W_0$ and rescaling
the resonant amplitude coefficients $b_k$, $c_k$, and $b$.

The only hypothesis that we have found to be in a statistically significant
agreement with the data is that the cross section at 4010 and 4015\,MeV for the
$D {\bar D}$ channel and at 4015\,MeV for the $D^* {\bar D}^*$ channel is
contributed by a new resonance. The resonance has to be quite narrow, with a
width of at most few MeV, in order to not affect the cross section at the rest
of the scan energies. Within this hypothesis the 4010 and 4015\,MeV data for
these channels are to be removed from the fit, and the remaining data are to be
fit within the previously described model. At this point we cannot assess from
the available data if the suggested resonance has any significant coupling to
the $D^* {\bar D}$ or $D_s {\bar D}_s$ channels. In the fit we have made the
simplest assumption that such coupling can be neglected and retained all the
data for the latter channels in the fit. In this way we find that the quality of
the fit to the described subset of data is quite acceptable: $\chi^2/N_{\rm
dof}=3.0/5$, with the central values of the resonance parameters:
$W_0=4019\,$MeV, $\Gamma_0=65\,$MeV and $w=9.85\,$MeV. The plots corresponding
to this fit are shown in Fig.~1.

\begin{figure}[ht]
  \begin{center}
    \leavevmode
    \epsfxsize=15cm
    \epsfbox{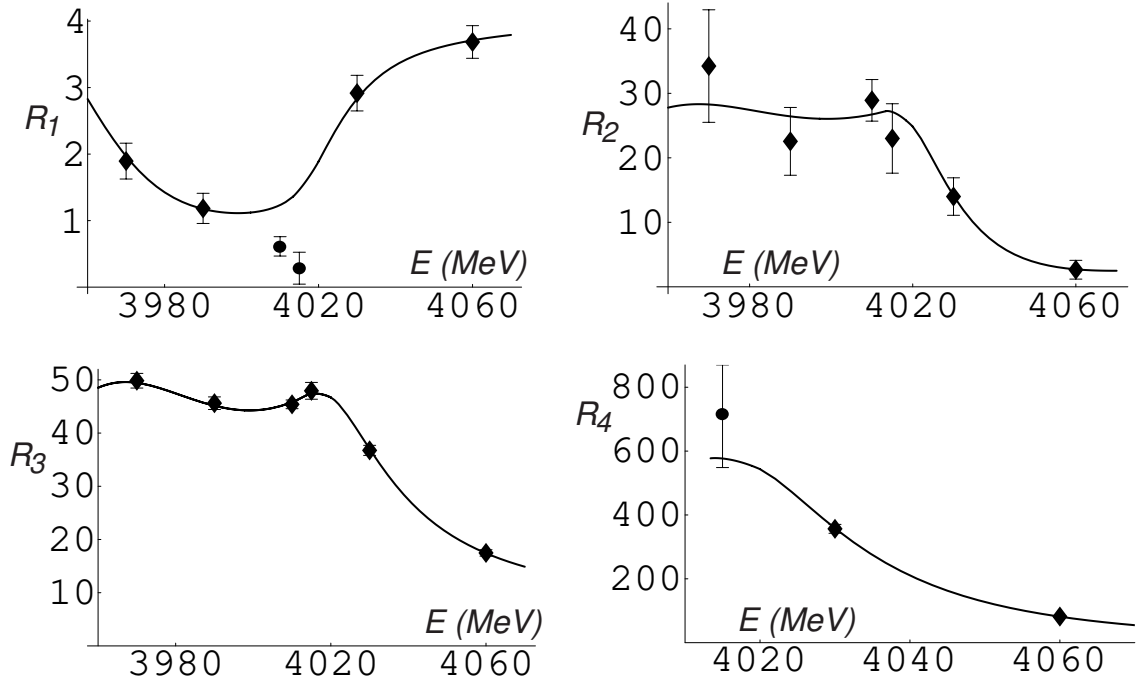}
    \caption{The plots of the rate coefficients $R$ corresponding to the fit
with excluded data points at energy 4010 and 4015\,MeV for the $D {\bar D}$
channel and at 4015\,MeV for the $D^* {\bar D}^*$ channel. The excluded points
are shown by filled circles.}
  \end{center}
\end{figure}

If the deviation of the data points at 4010 and 4015\,MeV indeed implies an
existence of relatively narrow resonance in addition to a broader and more
prominent resonance, conventionally associated with $\psi(4040)$~\footnote{We
use the PDG notation $\psi(4040)$ for the dominant resonance, even though its
`nominal' position is given at 4019\,MeV by the fit.}, an interesting problem
arises of explaining the co-existence of the two resonances with the same
quantum numbers $J^{PC}=1^{--}$ but apparently with strikingly different decay
properties.

One such difference, in the strength of the coupling to $e^+e^-$, can in fact be
estimated from the data. Indeed, if the production of $D^* {\bar D}^*$ at
$E=4030\,$MeV is dominated by the resonance $\psi(4040)$, the $e^+e^-$ width of
the resonance, $\Gamma_{ee}(4040)$, can be found from the well known formula
\beq
\Gamma_{ee}={W_0^2 \, \sigma(e^+e^- \to D^* {\bar D}^*) \over 3 \pi \,
\Gamma_{D^* {\bar D}^*}} \, \left [ (E-W_0)^2 + {1 \over 4} \, (\Gamma_0+
\Gamma_{D^* {\bar D}^*})^2 \right ]~,
\label{gee}
\eeq
where
\beq
\Gamma_{D^* {\bar D}^*}={1 \over 2} \left [ {(E-2 \, M_0)^{3/2} \over
w^{1/2} }+ {(E-2 \, M_+)^{3/2} \over w^{1/2} } \right ]
\label{gdd}
\eeq
is the (energy-dependent) width of the resonance decay into $D^* {\bar D}^*$
pairs. Using the results of the fit for the parameters $W_0$, $\Gamma_0$, and
$w$, one can readily estimate $\Gamma_{D^* {\bar D}^*}(4030) \approx 16\,$MeV,
and $\Gamma_{ee} \approx 1.7\,$KeV .

The coupling to the $e^+e^-$ channel of the suggested new resonance $X$ can be
very approximately estimated from a formula, similar to Eq.(\ref{gee}) applied
at the maximum of the resonance:
\beq
{\cal B}_{ee}(X)={ M_X^2 \, \sigma(e^+e^- \to X)_{\rm max} \over 12 \pi}~.
\label{bee}
\eeq
An estimate of the cross section, associated with the hypothetical new resonance
is naturally uncertain. The deviation of the data point at 4015\,MeV in the $D^*
{\bar D}^*$ channel is comparable to the error, and is about 0.15 of the already
small (due to very small phase space) cross section $\sigma(e^+e^- \to D^* {\bar
D}^*) \approx 0.15\,$nb. Therefore only about 0.02 of the $e^+e^-$ annihilation
cross section at this energy can be associated with the $D^{*0} {\bar D}^{*0}$
channel. On the other hand in the $D {\bar D}$ channel the cross section has a
minimum at 4015\,MeV, apparently due to a destructive interference with the
background. Thus, if the `would be cross section' $\sigma(e^+e^- \to X \to D
{\bar D})$ is estimated as the depth of this minimum (i.e. assuming maximal
negative interference):  $\sigma(e^+e^- \to X \to D {\bar D}) \sim 0.1 \,$nb,
one can arrive at the estimate ${\cal B}_{ee}(X) \sim 10^{-7}$. Since, as can be
seen from the plot for $R_1$ in Figure 1, the total width of the $X$ resonance
is likely to be few MeV, one can estimate the $e^+e^-$ width of $X$ in the
ballpark of few tenths of eV.

Clearly, such large ratio of the $e^+e^-$ widths (of order $10^4$) of two
closely spaced resonances implies that either the mixing between them is finely
tuned and results in a great cancellation of the coupling of $X$ to $e^+e^-$, or
the mixing is very small: not larger than $O(0.01)$ in the amplitude. A small
mixing can be due to a completely different structure of the two states. Indeed
the dominant relatively broad resonance can well be a $3^3S_1$ charmonium state.
The estimated $\Gamma_{ee}$ is close to that predicted for such a state in a
charmonium model\cite{cornell} ($\Gamma_{ee}=1.5\,$KeV), although the position
is somewhat off that in the model ($M(3S)=4.11\,$GeV). It can be also noted that
our estimate of $\Gamma_{ee}$ as well as the model prediction is somewhat higher
than the PDG value $\Gamma_{ee}=(0.75 \pm 0.15)\,$KeV for $\psi(4040)$. However,
given the uncertainties and that the data and their analyses are still in flux,
we believe that the disagreement is not dramatic. In either case it is quite
clear that the $e^+e^-$ width of $\psi(4040)$ is in the KeV range, i.e. that
typical of a $3^3S_1$ charmonium state. On the other hand, the new hypothetical
resonance $X$ can be naturally explained as a $D^* {\bar D}^*$ $P$-wave
molecule, similar to the $S$-wave $D^{*0} {\bar D}^0$ molecule X(3872). In
particular the small width $\Gamma_{ee}(X)$ then agrees with the
expected\cite{ov} $e^+e^-$ width of a $P$-wave molecule with typical size
$O(1\,{\rm fm})$.

The considered here interpretation of the peculiar behavior of the CLEO-c scan
data in the immediate vicinity of the $D^* {\bar D}^*$ threshold in terms of
existence of a narrow resonance $X$ in addition to $\psi(4040)$ can be either
confirmed or rejected if more detailed $e^+e^-$ data in this energy region
become available. We would like to point out in this connection that it would
also be extremely instructive if angular data for the $D^* {\bar D}^*$ channel
were available. Namely, as previously mentioned, the vector meson pairs can be
produced in two $P$-wave states with different total spin: $S=0$ and $S=2$. If
the picture of two resonances considered here is correct, then it is highly
likely that the states $X$ and $\psi(4040)$ are coupled to quite different
combinations of these two spin states. The interplay between the two resonances
should then result in a rapid variation with energy of a measurable angular
correlation, e.g. the angular distribution of the pions from the decays $D^* \to
D \pi$ with respect to the direction of the initial beams, which distribution is
sensitive to both the relative absolute value and the phase of the two spin
amplitudes\cite{mv2}.

While this work was finalized there appeared the paper \cite{belle06} presenting
the Belle data on the cross section of the $e^+e^-$ annihilation to the $D^*
{\bar D}^*$ and $D^* {\bar D}$ final states near and above 4.0\,GeV. This
encourages one to expect that more detailed results will become available soon,
including possibly the data on the $D {\bar D}$ annihilation channel.

This work is supported in part by the DOE grant DE-FG02-94ER40823.

\end{document}